\documentclass[
 aps,
 prl,
 amsmath,amssymb,
 reprint,%
author-numerical,%
superscript address,
 showpacs,
]{revtex4-1}

\usepackage{graphicx}
\usepackage{dcolumn}
\usepackage{bm}

\usepackage{color}
\usepackage{multirow}
\makeatletter

\newcommand*{\Rmnum}[1]{\expandafter\@slowromancap\romannumeral #1@}
\makeatother

\begin{document}


\title[prl]{
Cation ordering induced polarization enhancement for PbTiO$_3$-SrTiO$_3$ ferroelectric-dielectric superlattices
}

\author{Junkai Deng}
\affiliation
{State Key Laboratory for Mechanical Behavior of Materials, Xi'an Jiaotong University, Xi'an 710049, China}

\author{Alex Zunger}
\affiliation
{University of Colorado, Boulder, Colorado 80309, USA}

\author{Jefferson Zhe Liu}
\email {zhe.liu@monash.edu}
\affiliation
{Department of Mechanical and Aerospace Engineering, Monash University, Clayton, Victoria 3800, Australia}


\begin{abstract}
In this paper, an efficient computational material design approach (cluster expansion) is employed for the ferroelectric PbTiO$_3$/SrTiO$_3$ system. Via exploring a configuration space including over $3 \times 10^6$ candidates, two special cation ordered configurations: either perfect or mixed 1/1 (101) superlattice, are identified with the mostly enhanced ferroelectric polarization by up to about 100\% in comparison with the (001) superlattice. Analyzing these two special configurations reveals the exotic couplings between the antiferrodistortive distortion (AFD) and ferroelectric polarization (FE) modes, \emph{e.g.}, tilting along $x$ or $y$ axis and polarization in $z$ direction, as the origin for the best polarization property in this system. The identified cation ordering motifs and the exotic AFD-FE couplings could provide fresh ideas to design multifunctional perovskite heterostructures.

\end{abstract}

\maketitle



Ferroelectric (FE) thin films and superlattices (SLs) based on perovskite oxide are currently the subject of intensive research due to their promising properties for low power electronics, ultrastable solid-state memory, sensors, and medical imaging technologies, and also because of the fundamental scientific importance \cite{uchino2009ferroelectric}. For most perovskite ABO$_3$, the zone-center ferroelectric (FE) distortion, characterised by the opposite motion of the cations with respect to the oxygen cage, and the non-polar zone-boundary antiferrodistortive distortion (AFD) modes, which consist of rotation and titling of the oxygen octahedra surrounding the B cation or the anti-polar motion of A-site cations, often tend to suppress each other \cite{lines1977book}. This gives rise to a small number of perovskite ferroelectrics in nature \cite{Benedek2013}. 

Fortunately, the balance is extremely delicate and it can be tuned for the coexistence, by assembling different types of perovskite 5-atom units into different ordered configurations \cite{Wu2011PRB,Schlom2007ARMR,Kim2013AM,Rondinelli2012AM,Rondinelli2011AM}. For instance, the recently discovered hybrid improper ferroelectricity in ultra-short (001) perovskite superlattice possesses a new form of interface coupling of FE and AFD octahedral rotational distortion \cite{Benedek2011, Fukushima2011, Benedek2012, Young2013, Sim2013}, giving rise to FE polarization, despite no FE in the parent perovskites. Additionally, in ultra-short (001) PbTiO$_3$/SrTiO$_3$ (PTO/STO) superlattice, it is found that oxygen octahedra rotation promotes the polarization, leading to a surprising enhancement of FE polarization \cite{Bousquet2008Nature, Puente2011PRL}. The buckling of the inter-octahedra B-O-B bond angles, a direct consequence of the octahedra rotation, can change physical properties of the perovskite oxides, for example, electronic bandwidth, magnetic interactions, and critical transition temperatures and so on \cite{Imada1998RMP}. The strong coupling of the AFD and FE distortions, therefore, presents a great opportunity to create novel multifunctional materials that respond to external electric fields, for example, multiferroics \cite{Bousquet2008Nature, Ramesh2007}.

The design of perovskite heterostructures that possess a AFD/FE coupling, however, is nontrivial. Some elegant design criteria have been proposed recently \cite{Puente2011PRL, Wu2011PRB, Bousquet2008Nature, Rondinelli2012AM, Rondinelli2012MRS, Rondinelli2011AM, Mulder2013AFM}. But they only considered a limited number of configurations, such as the 1/1 and 2/2 (001) SLs etc.. In addition, the proposed design rules try to relate properties of the SLs to properties of the parent single-phase perovskite, aiming to predict the properties of a particular SL by just considering the properties of the constituent phases. But the vast possibilities of A- or B-site ordered configurations in perovskite heterostructures and different chemistry of the cations could in-principle give rise to exotic properties that do not exist in the parent constituents. It is thus highly desirable to develop an efficient computational method to explore all possible configurations for the design of perovskite heterostructures with the best properties.

The cluster expansion (CE) approach \cite{fontaine1994, Laks} can map the relations between different configurations of building-units on a given Bravais lattice and their physical properties. For a binary mixture, one defines a configuration $\sigma$ as a specific decoration of two types of building-units on a lattice, in which each lattice site is occupied by either of the two (spin variable $s_{i}=-1$ or $=1$, respectively). The property of interest $\mathcal{F}$ can then be expressed as:
\begin{equation}
   \mathcal{F}_{\text{CE}}(\sigma)=J_{0}+\frac{1}{N} [\sum_{i}{J_{i}s_{i}} + \sum_{i,j}{J_{ij}s_{i}s_{j}} + \sum_{i,j,k}{J_{ijk}s_{i}s_{j}s_{k}} + \cdots]
   \label{ce}
\end{equation}
where $J_{ij}$, $J_{ijk}$, $\cdots$ represent the effective-cluster interactions (ECIs) for pair, three-body, $\cdots$, interactions in the chemical system, and the $s_{i}s_{j}$, $s_{i}s_{j}s_{k}$, $\cdots$ are the multisite cluster functions that form a complete basis set in the configuration space \cite{Sanchez1984PhysicaA}. The ECIs can be obtained by fitting the first-principles calculated results ($\mathcal{F}$) of a set of ordered configurations to Eq. (1). In principle, the CE method can be applied to any physical property if it is a well-defined functional of configurations, for example, total energy \cite{fontaine1994}, Curie temperature \cite{Franceschetti}, elastic modulus \cite{LiuZ05}, thermo-conductivity \cite{Chan}, and so on. 


In this letter, we adopted the CE approach for FE polarization $P$ of the PTO/STO system that involves two types of 5-atom building-units (PTO and STO) decorated on a simple cubic lattice (\textit{i.e.}, sub-lattice A of perovskite). The $P_{\text{CE}}(\sigma)$ functional was fitted using the first-principles results of a handful of configurations and then was used to explore a huge configuration space 
to search candidates with the best FE polarization. Through analysing the two identified configurations, the underlying physical mechanism was disclosed. Note that in some high-throughput methods one attempts to calculate in principle all structures/configurations in the selected space  (\textit{e.g}, ICSD) \cite{MatProj, AFLOW, AFLOWLib}, here we use a low-throughput first-principles method to calculate but O(50) structures, then parametrize a surrogate model (\textit{i.e.}, CE) from which we obtain essentially effortlessly the results with comparable accuracy for O($3\times10^6$) structures. Once the latter configuration space is searched, one can return to direct first-principles calculations, however, aimed only at the O(10) `best of class' configurations identified as the most promising. This represents an enormous saving relative to the ordinary high-throughput approach of calculating all configurations at the outset.

First-principles calculations based on density functional theory (DFT) were performed using the local density approximation \cite{LDA} and the projector augmented wave method \cite{PAW} implemented in the Vienna Ab Initio Simulation Package \cite{Kresse1996PRB}. The in-plane lattice constant was fixed to 3.864 {\AA} to account the constraint from a cubic STO substrate. The use of periodic boundary conditions imposed short-circuit electrical conditions. All ionic positions were relaxed until the forces were less than 5 meV/{\AA}. The obtained cation off-centre displacements and the bulk Born effective charges, \textit{i.e.}, $Z^*_{\text{Pb,Sr}}=2.7$ and $Z^*_{\text{Ti}}=4.6$, were used to compute the electric polarization results $P_{\text{DFT}}(\sigma)$. This method yielded an excellent agreement with results using the Berry phase method \cite{Cooper2007PRB}.

An iterative training process was used to fit the ECIs in Eq. (1). We started with feeding the $P_{\text{DFT}}(\sigma)$ results of 23 usual suspects to fit the ECIs. Using the obtained $P_{\text{CE}}(\sigma)$ functional, an exhaustive enumeration method was employed to search a configurational space (O(3 $\times 10^6$) configurations) for the largest polarization (LP) configurations. The $P_{\text{DFT}}$ results of these LP candidates, if not available, were then determined using DFT calculations and compared against the CE predictions. In the next iteration, these $P_{\text{DFT}}$ results were added to the DFT data pool to refit the ECIs, and then the obtained $P_{\text{CE}}(\sigma)$ was used to search for the new LP candidates. This iterative process was repeated until the CE predicted results agreed with the DFT calculations and no new LP configurations were predicted. A good convergence was achieved with only 48 DFT inputs. The obtained $P_{\text{CE}}(\sigma)$ includes 15 pairs, 1 triplet, 3 quadruplet, and 1 quintuplet clusters. The cross-validation score \cite{vandeWalle}, representing the prediction error of $P_{\text{CE}}(\sigma)$, is less than 0.011 C/m$^2$.

\begin{figure}[!htb]
\includegraphics[width=85mm]{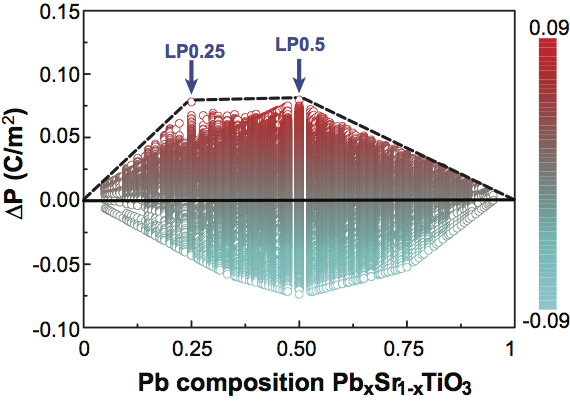}
\caption{\label{Fig.1} Polarization enhancement of PTO-STO configurations with respect to the concentration weighted average of bulk PTO and STO, $\Delta P_{\text{CE}}(\sigma)$, predicted by the cluster expansion approach. The two configurations at breaking points of the convex hull represent the optimal configurations with the largest polarization (LP) at PTO concentration of 0.25 and 0.5 (named LP0.25 and LP0.5), respectively.} 
\end{figure}


\emph{The largest polarization configurations} Figure 1 and Fig. S1 show the polarization enhancement $\Delta P_{\text{CE}}(\sigma)$ of the O($3\times 10^6$) ordered configurations with respect to the concentration weighted average of the bulk PTO and STO. Two configurations at breaking points of the convex hull (PTO concentration of 0.25 and 0.5) are the LP structures that exhibit the mostly enhanced FE polarization. Their crystal structures are presented in Fig. 3 and Fig. S2. Their $P$ values are 0.237 C/m$^2$ and 0.392 C/m$^2$, \textit{i.e.}, an enhancement of 95\%, and 71\% in comparison with the counterpart (001) SLs, respectively (Fig. S1). 

Careful inspection reveals a common cation ordering motif in Fig. 3 and Fig. S2: perfect or mixed (101) SLs. The LP0.5 is a perfect (PbTiO$_3$)$_1$/(SrTiO$_3$)$_1$ (101) SL. The LP0.25 is an intermixed (Pb$_{0.5}$Sr$_{0.5}$TiO$_3$)$_1$/(SrTiO$_3$)$_1$ (101) SL, where in the intermixed (101) plane, the Pb and Sr cations are located next to each other in a column along the [100] axis, and the two adjacent [100] columns appear a relative shift in [100] direction. Such special configurations were not expected before but successfully identified by CE. The reason why this special motif results in the significantly enhanced polarization will be discussed later.

\begin{figure}[!htb]
\includegraphics[width=85mm]{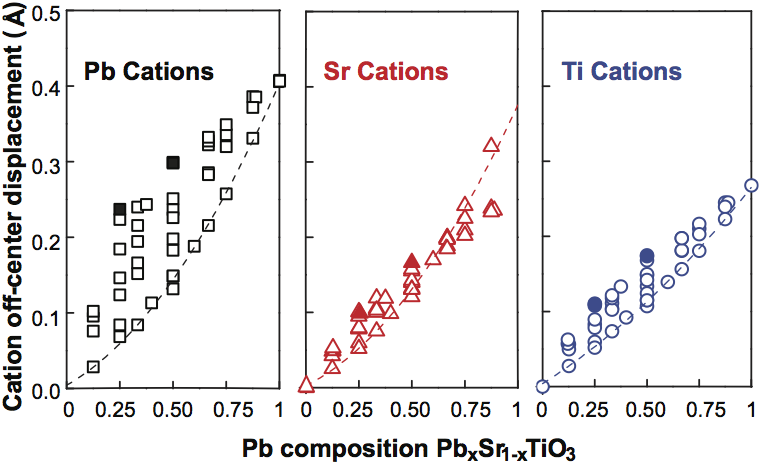}
\caption{\label{Fig.2}Average off-center displacements of (a) Pb, (b) Sr and (c) Ti cations of the DFT calculated 48 configurations. The dashed lines are fitted to off-center displacements of the (001) superlattices.}
\end{figure}

\begin{figure*}[!htb]
\includegraphics[width=150mm]{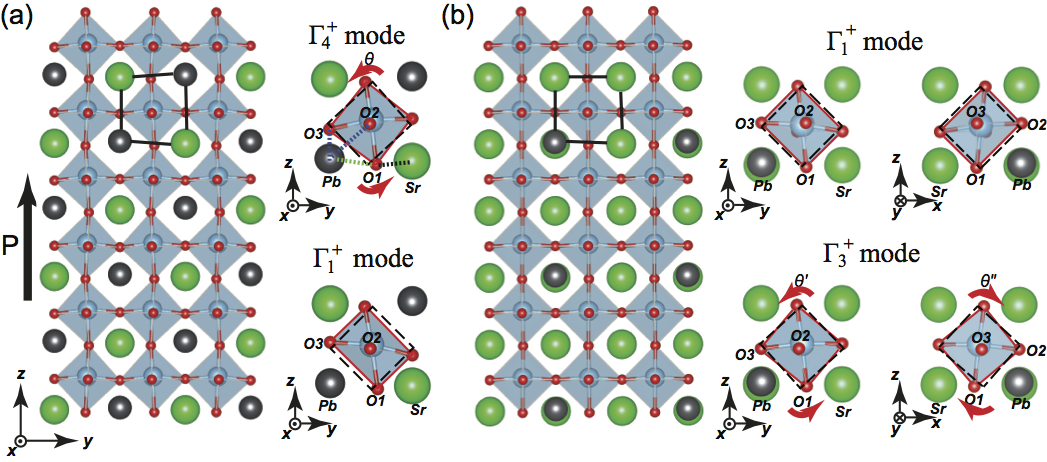}
\caption{\label{Fig.3}Relaxed structures and two primary AFD distortion modes of the (a) LP0.5 and (b) LP0.25 in DFT calculations. In LP0.5, the primary mode $\Gamma_4^+$ represents an octahedra tilting along $x$ axis with oxygen atom in (001) plane moving toward Sr cations (the Glazer notation $a^+b^0b^0$). The secondary $\Gamma_1^+$ mode represents a shape distortion of octahedra with oxygen atoms O$_1$ and O$_3$ moving toward Sr cations. Interestingly the two primary modes of LP0.25, $\Gamma_1^+$ and $\Gamma_3^+$ (Glazer notation $a^+a^+c^0$), resemble those of LP0.5. See text for details.}
\end{figure*}

\emph{Coupling of AFD distortion and FE polarization in the LP configurations} The roles of different cations in determining the FE polarization can be seen in Fig. 2, where averaged cation off-centre displacements of the 48 configurations calculated using DFT are summarised. The dashed lines in Fig. 2 depict results for the (001) PTO/STO superlattices. The solid symbols represent results for those two LP configurations (Fig. 1), indicating the largest cation off-centre displacements (thus the largest FE polarization). A much less variation of the averaged Sr and Ti off-centre displacements with respect to other configurations suggests the primary role of the Pb cations.

To elucidate origin of the enhanced polarization of the LP configurations, the relaxed crystal structures of LP0.5 and LP0.25 (Fig. 3) are carefully examined using the symmetry mode analysis program ISOTROPY suite \cite{Campbell}. Two primary AFD distortion modes of the LP0.5 are shown in Fig. 3(a). The {$\Gamma_4^+$} mode represents a tilting of octahedra about $x$ axis with a titling angle of $\theta = 2.9^\text{o}$, consequently in (001) plane the oxygen atoms moving away from the Pb and toward the Sr cations. The {$\Gamma_1^+$} mode represents a shape distortion of octahedra, \emph{i.e.}, the Ti-O bonds being bent toward the Sr cations in the $yz$ plane. It is interesting to note that the two primary AFD modes of the LP0.25 [Fig. 3(b)], {$\Gamma_1^+$} and {$\Gamma_3^+$}, resemble those of LP0.5, except that in mode {$\Gamma_1^+$} the Ti-O bonds are bent in both the $xz$ and $yz$ planes and the octahedra tilting in {$\Gamma_3^+$} mode takes place along both $x$ and $y$ axis. 

To examine the coupling between the AFD modes and the FE modes, we artificially "turn off" these modes in the LP structures meanwhile allowing the FE polarization to fully develop in DFT calculations. The results are shown in Table \ref{Tab1}. Turning-off the minor distortion modes ('others' in Table \ref{Tab1}, more information in Table SI) in LP0.5 slightly enhance the $P$ to 0.396 C/m$^2$. It suggests that these AFD distortion modes indeed suppress the FE polarization, consistent with the traditional views in perovskites. In contrast, {$\Gamma_4^+$} and {$\Gamma_1^+$} modes promote the polarization, turning-off of these two modes significantly reducing the polarization value down to 0.354 C/m$^2$ together with an increase of total energy. The same conclusion can be drawn for LP0.25, \emph{i.e.}, the AFD modes enhancing the $P$ values monotonously from 0.209 to 0.237C/m$^2$. Note that for the PTO/STO 1/1 or 2/2 (001) SLs, the AFD$_z$ and FE$_z$ coupling \cite{Bousquet2008Nature} and the coupling between in-plane titling AFD$_{xy}$ and the in-plane polarization FE$_{xy}$ \cite{Puente2011PRL} have been observed. The discovered coupling between in-plane AFD$_{x}$ or AFD$_{xy}$ (\emph{i.e.}, $\Gamma_4^+$ in LP0.5 and $\Gamma_3^+$ in LP0.25) and out-of-plane FE$_z$ in our LP structures has not been reported. These distortion modes and their couplings do not exist in the parent constituents, either. 

Note that the LP configurations can be seen as intermixed (001) SLs (Fig. \ref{Fig.3} and Fig. S2). It was proposed that by introducing some degree of interface cation mixing to a ferroelectric-dielectric superlattice could enhance FE polarization \cite{Cooper2007PRB}. This is because to avoid the electrostatic energy penalty from the net charge accumulated at interfaces, the FE/dielectric layer will be depolarized/polarized to yield similar polarization values, often leading to a reduced total polarization value \cite{Bousquet2008Nature, Cooper2007PRB}. Introducing interlayer cation mixing could reduce the energy penalty and thus enhance the polarization $P$. In Table \ref{Tab1}, the intermixing effect of LP0.5 indeed reduces the total energy and increases the polarization significantly in comparison with the (001) SL, apparently supporting the model from Cooper et al.. \cite{Cooper2007PRB} But an exception is observed for LP0.25. The intermixing does not reduce the energy penalty, whereas it significantly enhances the $P$ value. It appears that there are no simple theoretical models to account for the interface mixing effect on the FE polarization.  

The AFD/FE coupling plays a decisive role in determining the LP structures. Our DFT calculations show that a Sr cation dopant in PbTiO$_3$ tends to attract oxygen atoms moving toward it (Fig. S4) \cite{Puente2011PRL}. For the LP0.5, the check-board cation ordering pattern in the $yz$ plane [Fig. 3(a) and Fig. S2] ensures the two Sr cations (neighbouring a Pb cation) can work collaboratively to maximise the $\Gamma_4^+$ and $\Gamma_1^+$ modes. In the LP0.25, the Pb cations show a BCC ordered pattern on the simple cubic Bravis lattice (Fig. S2), in which all 6 nearest neighbours of every Pb cation are the Sr cations. These Sr neighbours should work collaboratively to maximise the $\Gamma_1^+$ and $\Gamma_4^+$ modes. Structural analysis indicates that the LP0.5 and LP0.25 have the largest octahedra titling angle and Ti-O-Ti bending angle among all the configurations studied in our DFT calculations at $x=0.25$ and $0.50$ (Fig. S5 and Table SIII), respectively. For a comparison, we noticed one specific structure str381 (Fig. S6), which has a very similar cation ordered pattern as LP0.5. If only considering the intermixing effect, the str381 has a lower total energy and a higher $P$ value than those of LP0.5 (Table SIV). But the small variation in ordered pattern results in a weaker collaborative effect, leading to a smaller tilting angle and Ti-O bending angle (Table SIII). The AFD/FE coupling thus yields a larger $P$ increase for the LP0.5 to overtakes str381 as the LP configuration at $x=0.5$ (Table SIV). We believe that the AFD/FE coupling (\emph{e.g.}, AFD$_x$ or AFD$_{xy}$ and FE$_z$) is the origin for LP structures to have the largest polarization enhancement among all the $3\times 10^6$ candidates.
 
\begin{table}[!b]
\centering
\caption{\label{Tab1} FE polarization and relative total energy results of the LP0.5 and LP0.25 with the distortion modes subsequently turning-off. A comparison is made with the perfect (001) SLs. In all the cases, the polarization is allowed to fully develop in DFT calculations.}
\begin{tabular}{c l c c c c c c}
\toprule
    & modes & & P$_z$ & & $\Delta E_{rel}$  \\
    & & & (C/m$^2$) & & (meV/cell) \\
\hline
  \multirow{4}{*}{LP0.5} & $\Gamma_4^+$ {\&} $\Gamma_1^+$ {\&} Other & & 0.392 & & -15.36 \\
                         & $\Gamma_4^+$ {\&} $\Gamma_1^+$ & & 0.396 & & -15.11 \\               
                         & $\Gamma_4^+$ & & 0.381 & & -5.33 \\                        
                         & intermixing & & 0.354 & & -3.65 \\

\hline
   \multirow{1}{*}{(001) 1/1 SL} &   & & 0.229 & & 0.0\\
\toprule
  \multirow{4}{*}{LP0.25} & $\Gamma_1^+$ {\&} $\Gamma_3^+$ {\&} Other & & 0.237 & & -4.0 \\
                         & $\Gamma_1^+$ {\&} $\Gamma_3^+$ & & 0.232 & & -3.95 \\                         & $\Gamma_1^+$ & & 0.224 & & -3.54 \\                      
                         & intermixing & & 0.209 & & 2.5 \\
\hline
   \multirow{1}{*}{(001) 1/3 SL} &   & & 0.122 & & 0.0\\
\toprule
\end{tabular}
\end{table}

\begin{figure}[!htb]
\includegraphics[width=61mm]{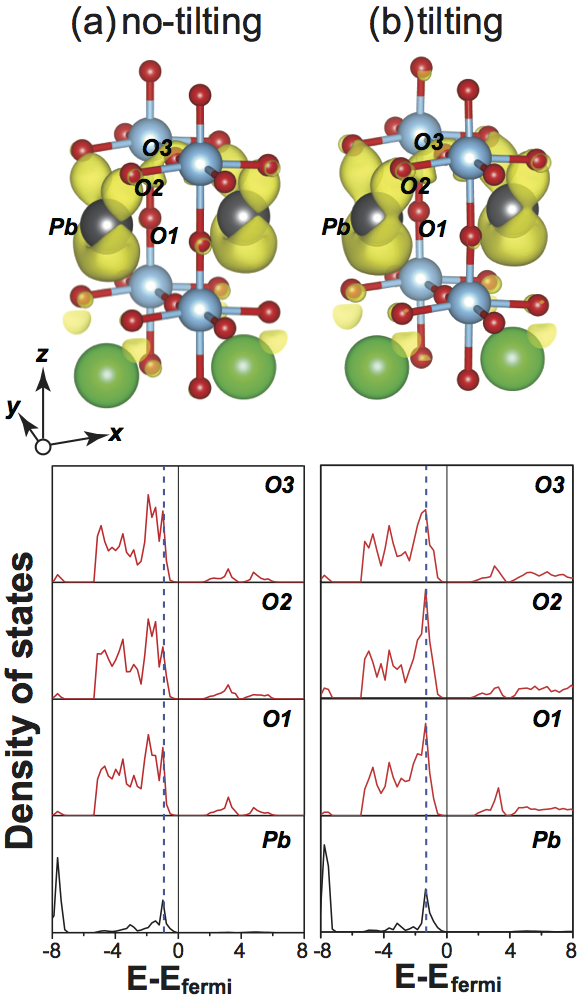}
\caption{\label{Fig.4} Difference-electron density and partial density of states (pDOS) for Pb and neighboring O ions in LP0.5 without octahedral tilts (a) and with titling (b). The yellow iso-surfaces are plotted at 0.0045 $e^-/$bohr$^3$. In partial DOS results, black lines represent Pb 6s states and red lines represent O 2p states.}
\end{figure}

\emph{Physical origins for the AFD$_{x}$/FE$_z$ coupling} Figure 4 shows the difference-electron density analysis and the projected partial density of states (pDOS) for Pb and its neighbouring O ions in the LP0.5 (without/with AFD$_x$). Clearly, the octahedra titling can significantly enhance the hybridisation between the Pb 6s  and the O 2p orbitals, evidenced by the strongly overlapped peaks in pDOS and the strong charge redistribution among the Pb and O ions. For bulk PTO, it is well established that the hybridisation between the Pb lone pair 6s electrons and the O 2p electrons induces its superior FE properties. The enhanced electronic hybridisation caused by the AFD$_x$ titling is, therefore, origin for the observed increase of FE$_z$ polarization \cite{Puente2011PRL}.

Despite the recent progresses \cite{Rondinelli2012MRS, Rondinelli2011AM}, developing design rules for perovskite heterostructures is a complex issue. Our study is an example case, where the exotic and unexpected couplings induces the best FE polarization in the PTO/STO system. This surprising result indicates the rich and novel physics embodied in pervoskite systems, which remain to be explored. Our CE method could serve as a general means to explore the huge configuration space ($3 \times 10^6$ candidates) for discoveries. 

In summary, an efficient computational material design approach (cluster expansion) was used to study the cation ordering effects on the FE polarization in PTO/STO perovskite. Two (101) superlattices are identified as the configurations with the best polarization property. The exotic AFD/FE couplings (\emph{e.g.}, between in-plane titling AFD$_x$ or AFD$_{xy}$ and out-of-plane polarization FE$_z$) are revealed as the origin for the best polarization $P$ in this system. The identified LP motifs and the exotic coupling could provide fresh ideas for the design of perovskite heterostructures.

\begin{acknowledgments}
JD and JZL acknowledge the financial support from Monash University Engineering faculty 2013 seed grant. JD also thanks the support of NSFC (51320105014, 51471126), China Postdoctoral Science Foundation (2014M552435). Work of AZ was funded by DOE, office of science, Basic energy science. The authors gratefully acknowledge computational support from Monash University Sun Grid and the National Computing Infrastructure funded by the Australian Government.
\end{acknowledgments}

\bibliography{references}


\end{document}



\title[prl]{\textit{Supplemental materials for the manuscript}\\
\vspace{1cm}
Cation ordering induced polarization enhancement for PbTiO$_3$-SrTiO$_3$ ferroelectric-dielectric superlattices}


\author{Junkai Deng}
\affiliation
{State Key Laboratory for Mechanical Behavior of Materials, Xi'an Jiaotong University, Xi'an 710049, China}

\author{Alex Zunger}
\affiliation
{University of Colorado, Boulder, Colorado 80309, USA}

\author{Jefferson Zhe Liu}
\email {zhe.liu@monash.edu}
\affiliation
{Department of Mechanical and Aerospace Engineering, Monash University, Clayton, Victoria 3800, Australia}




\maketitle

\section{First-principles calculations}
DFT calculations were performed using the local density approximation (LDA) \cite{LDA} and the projector augmented wave (PAW) method \cite{PAW} implemented in the Vienna Ab Initio Simulation Package (VASP) \cite{Kresse1996PRB}. A plane wave energy cutoff of 600 eV was adopted and a Monkhorst-Pack k-point mesh of $12\times 12\times 12$ in a five atom perovskite unit cell was used for the Brillouin zone integration. For other cation ordered configurations (i.e., superstructures based on the 5 atom building unit), the same k-point mesh density was used. The constraint from a STO substrate is considered by fixing the in-plane lattice constant to that of the cubic STO (the stable structure at room temperature) 3.864 {\AA}. All ionic positions were relaxed until the forces were less than 5 meV/{\AA}. The use of periodic boundary conditions imposes short-circuit electrical conditions across the whole unit cell. The electric polarization was computed by using the bulk Born effective charges ($Z^*$), i.e., $Z^*_{\text{Pb,Sr}}=2.7$ and $Z^*_{\text{Ti}}=4.6$ and the cation off-centre displacement of a fully relaxed configuration in the LDA calculations \cite{Cooper2007PRB}. This method has been proved to lead to an excellent agreement with the results using the Berry phase method \cite{Cooper2007PRB}. The polarization of bulk PTO constrained to the STO substrate was calculated as 0.617 C/m$^2$.

Finding the lowest energy crystal geometry of a perovskite heterostructure (configuration) is a challenging task. We employed the following procedure to do the relaxation in our DFT calculations. 

(1) We create the ideal crystal structure from our configuration database (including $3 \times 10^6$ cation ordered configurations). (2) We introduce different distortion modes, such as ferroelectric polarisation, the in-phase and out-of-phase octahedral rotation about $z$ axis and titling about $x$ or $y$ axis, in the ideal structure, leading to several different distorted structures. Note that to accommodate these rotation/tilting distortions, the supercell size (in some cases) should be increased. (3) At a given c/a ratio, atomic positions in one distorted structure are fully relaxed. Fitting the obtained total energy results as a function of different $c/a$ ratios yields the lowest energy value and the corresponding $c/a$ ratio. (4) At this $c/a$ ratio, DFT calculation was carried out to relax atomic positions in the supercell. Then we calculated the off-center displacements of Pb, Sr, and Ti cations and then use the Born point charge model to determine the $P$ value. For each distorted structure, we repeat step (3) to (4). The one with the lowest total energy value is taken as the fully relaxed structure of the heterostructure (configuration).

\section{Polarization enhancement of PTO-STO configurations}

\begin{figure*}[!htb]
\includegraphics[width=90mm]{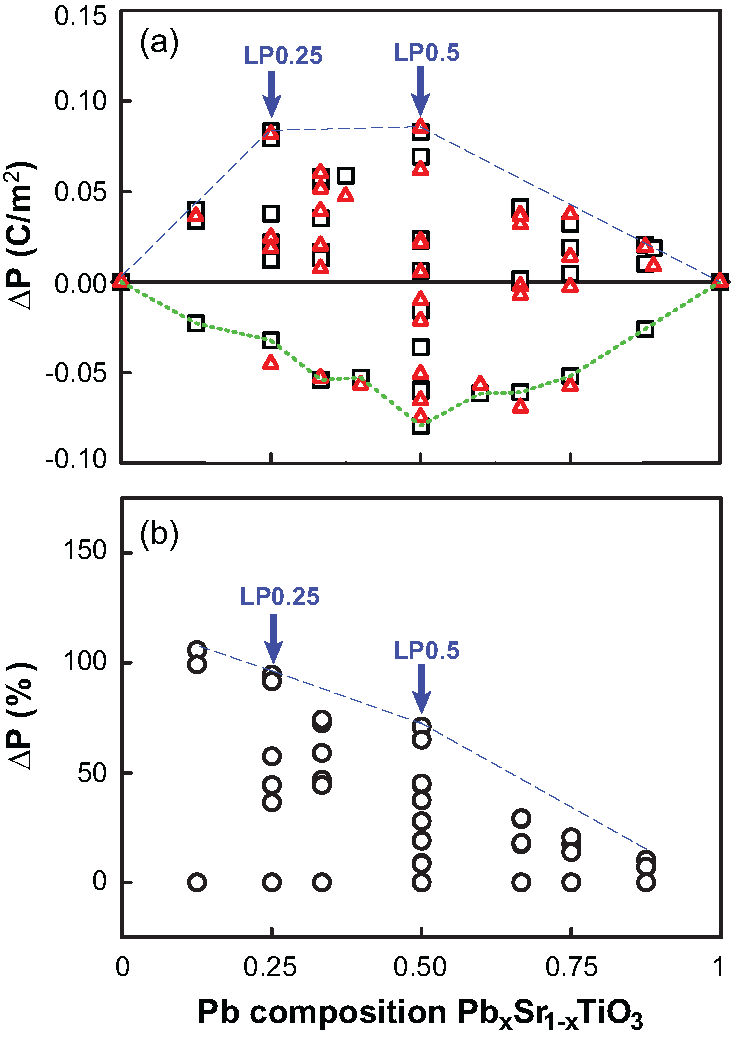}
\caption{\label{Fig-S1} (a) Polarization enhancement $\Delta P_{\text{CE}}(\sigma)$ of PTO-STO hetero-structures with respect to concentration weighted average of the bulk paraelectric STO and ferroelectric PTO. The black square and red triangle symbols represent the results obtained using DFT calculations and the predictions from the cluster expansion method, respectively. The green dashed line denotes the $\Delta P_{\text{DFT}}(\sigma)$ of the (001) superlattices. The two configurations at the corners of the convex hull (at PTO concentration of  0.25 and 0.50) represent the cation ordered configurations with the mostly enhanced polarisation. (b) The enhancement percentage of polarization for cation ordered structures with reference to the polarization results of the (001) superlattices.}
\end{figure*}

Polarization enhancement $\Delta P_{\text{LDA}}(\sigma)$ of a PTO-STO cation ordered configuration is defined as the difference of polarisation result with respect to the concentration weighted average of bulk STO and PTO. Figure S\ref{Fig-S1} shows the the results calculated using LDA $\Delta P_{\text{LDA}}(\sigma)$ (black square symbols) and compares with the predictions using the cluster expansion method $\Delta P_{\text{CE}}(\sigma)$ (red triangle symbols). The green dashed line in top figure denotes the results for (001) superlattices. The blue dashed line is the convex hull, where the two ordered configurations at the breaking points represent the configurations with the largest polarizations (LP) at 25\% and 50\% PTO composition. The bottom plot illustrates the percentages of polarization enhancement of a configuration with respect to the (001) superlattices at the same concentration. Those two LP configurations exhibit a remarkably enhanced the polarizations by 95\% and 71\%, respectively.

\begin{figure*}[!htb]
\includegraphics[width=130mm]{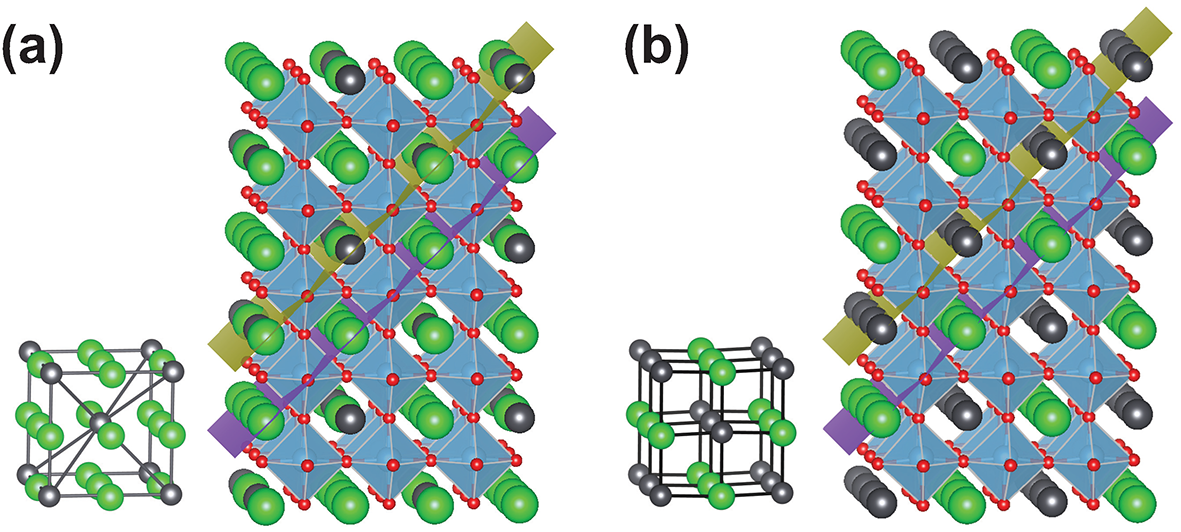}
\caption{\label{Fig-S2} 
Relaxed crystal structures of the (a) LP0.25, (b) LP0.5. They are either intermixed or perfect 1/1 (101) superlattices: (Pb$_{0.5}$Sr$_{0.5}$TiO$_3$)$_1$/(SrTiO$_3$)$_1$ or (PbTiO$_3$)$_1$/(SrTiO$_3$)$_1$. In LP0.25, the distribution of Pb cations exhibits a BCC pattern.}
\end{figure*}

Figure S\ref{Fig-S2} illustrates the relaxed crystal structures of the LP configurations at 25\% and 50\% PTO composition, respectively. These two structures share a common cation ordering motif: perfect or mixed (101) SLs. The LP0.5 is a perfect (PbTiO$_3$)$_1$/(SrTiO$_3$)$_1$ (101) SL. The LP0.25 is an intermixed (Pb$_{0.5}$Sr$_{0.5}$TiO$_3$)$_1$/(SrTiO$_3$)$_1$ (101) SL, where in the intermixed (101) plane, the Pb and Sr cations are located next to each other in a column along the [100] axis, and the two adjacent [100] columns has a relative shift in [100] direction. The Pb cations in LP0.25 form a BCC superstructure, in which all six nearest neighbours of every Pb cation are Sr cations. The group symmetry and Wyckoff positions of the fully relaxed crystal structures of LP0.5 and LP0.25 are presented in Appendix.

\section{Coupling between AFD and FE modes}

The symmetry mode analysis program ISOTROPY suite was employed to analyze the distortion modes of perovskite heterostructures \cite{Campbell}. The results for the LP0.5 and LP0.25 are summarized in Table S\ref{Tab-S1}. The space group for the ideal LP0.5 and LP0.25 crystal structure is P4/mmm and Im-3m, respectively. Different distortion modes lead to different space groups as shown in Table S\ref{Tab-S1}. The distortion modes are sorted in a descent order in terms of their magnitude. Some other minor distortion modes are neglected. Descriptions are provided for each distortion mode. 

\begin{table}[!ht]
\caption{\label{Tab-S1}Symmetry mode analysis for the LP0.5 and LP0.25 using program ISOTROPY suite. The group symmetry in parenthesis represents the symmetry of ideal crystal structure of LP0.5 and LP0.25 without any distortions. Glazer notation for some octahedra/titling modes is also included.}
\centering
\begin{tabular}{ c l l l }
\toprule
   {Structure} & {Distortion mode} & {Space group} & {Description} \\
\hline
LP0.5  & $\Gamma_5^-$ & 6 Pm & Ferroelectric polarisation \\
(123 P4/mmm)    & $\Gamma_4^+ (a^+b^0b^0)$ & 65 Cmmm & Octahedra tilting about $x$ axis \\
                            & $\Gamma_1^+$ & 123 P/mmm & Ti-O bond bending in $yz$ plane\\
                            & $Z_5^- (a^0a^0c^-)$& 11 P2-1/m & Octahedra rotation about $z$ axis, out-of-phase \\
                            & $Z_4^+$ & 132 P4-2/mcm & In (001) plane, two opposite oxygen corners of \\
                            & & & the octahedra moving align +z direction while \\
                            & & & the other two moving downwards \\
\hline        
LP0.25  & $\Gamma_4^-$ & 8 Cm & Ferroelectric polarisation \\
(229 Im-3m)         & $\Gamma_1^+$ & 229 Im-3m & Ti-O bond bending in $yz$ and $xz$ plane \\
                            & $\Gamma_3^+ (a^+a^+c^0)$ & 139 I4/mmm & Octahedra tilting about $x$ and $y$ axis\\
                            & $\Gamma_5^-$ & 42 Fmm2 & Ti anti-FE motion along (101) direction \\
\toprule
\end{tabular}
\end{table}

\begin{figure*}[!htb]
\includegraphics[width=120mm]{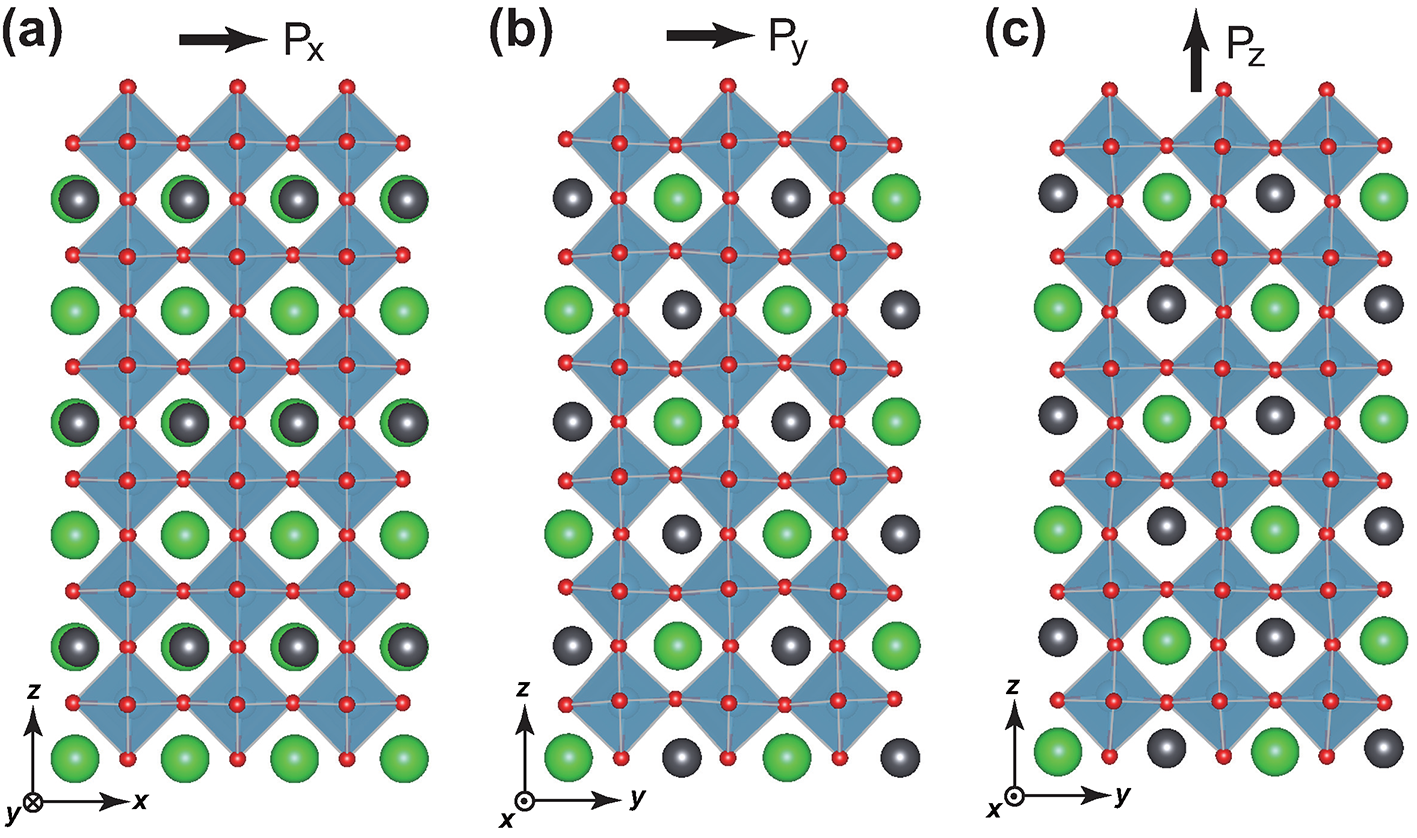}
\caption{\label{Fig-S3}Other possible AFD/FE couplings in the LP0.5 configuration. The relaxed structure of LP0.5 with (a) a coupling of FE$_x$ and AFD$_{xy}$, (b) a coupling of FE$_y$ and AFD$_{xy}$ and (c) a coupling of FE$_z$ and AFD$_{xy}$.}
\end{figure*}

For the LP0.5 configuration, other types of couplings between AFD octahedra rotation and FE polarization are also examined in LDA calculations, such as AFD$_{z}$/FE$_{z}$ coupling ($a^0a^0c^-$ and FE$_z$) \cite{Bousquet2008Nature}, AFD$_{xy}$/FE$_{xy}$ coupling ($a^+b^+c^0$ and FE$_{xy}$) \cite{Puente2011PRL}, and the AFD$_{z}$/FE$_{xy}$ coupling ($a^0a^0c^-$ and FE$_{xy}$). The results are summarised in Fig. S\ref{Fig-S3} and Table S\ref{Tab-S2}. It is clear that these coupling always leads to a higher total energy and a reduced polarisation value. We found that the AFD$_{z}$/FE$_{xy}$ coupling was not stable in LDA calculations. It is clear that the AFD$_x$/FE$_z$ coupling ($a^+b^0b^0$ and FE$_z$) leads to the lowest total energy and a highest $P$ result for the LP0.5.

\begin{table}[!htb]
\caption{\label{Tab-S2}LDA results for LP0.5 structure with different possible AFD/FE couplings as shown in Fig.S3.}
\centering
\begin{tabular}{c  c c  c c}
\toprule
   & {AFD$_{xy}$/FE$_x$} & {AFD$_{xy}$/FE$_y$} & AFD$_x$/{FE$_z$} & AFD$_z$/FE$_z$\\
\hline
 {polarization (C/m$^2$)}   & 0.207   & 0.267 & 0.392 & 0.231 \\
 {tilting/rotation angle (degree)}   & 0.29     & 1.30 & 2.90 & 4.55  \\
 {c/a ratio}                & 1.009   & 1.009 & 1.021 & 1.040 \\
 {Energy (eV/Unit-cell)}    & -42.957 & -42.964 & -42.975 & -42.966 \\
\toprule
\end{tabular}
\end{table}

\section{Insights obtained from analyzing the LP configurations}

\begin{figure*}[!htb]
\includegraphics[width=120mm]{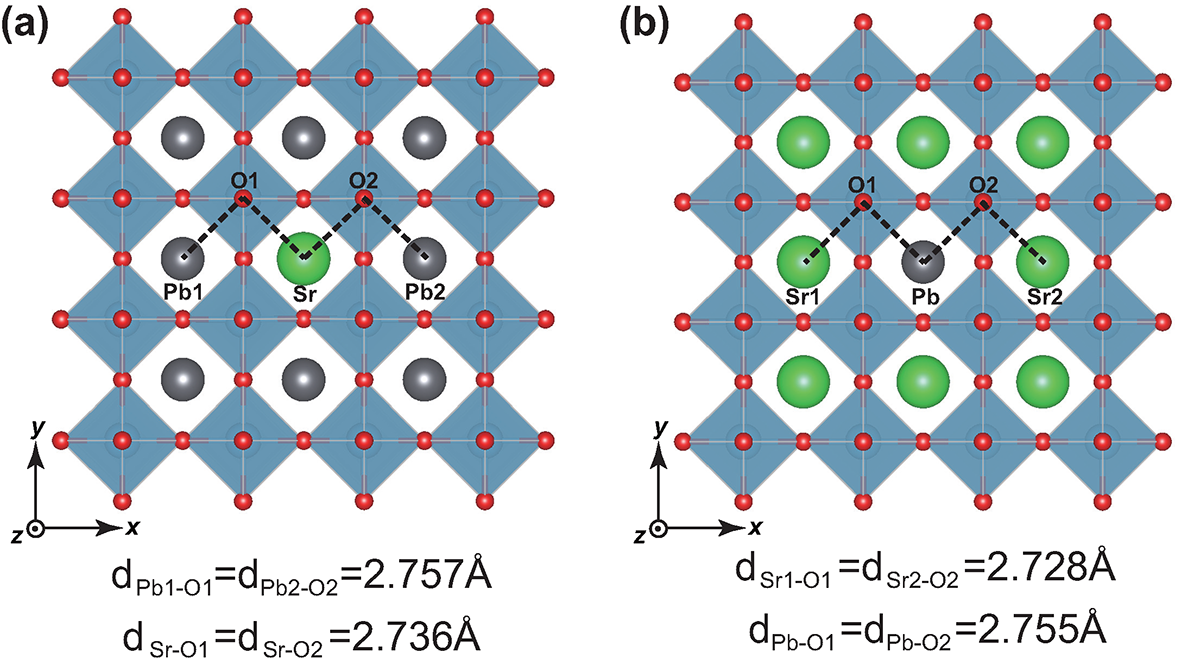}
\caption{\label{Fig-S4}The relaxed crystal structure of (a) bulk PTO with a dopant Sr cation and (b) bulk STO with a dopant Pb cation. The interatomic distances are also shown. The octahedra rotates so that the oxygen atoms move toward Sr cation in the (001) plane.}
\end{figure*}

Figure S\ref{Fig-S4} provides the evidences for different interactions of Pb and Sr with their neighbouring O atoms. When a Pb cation is replaced by Sr cation in a tetragonal ferroelectric PTO supercell ($4 \times 4\times 4$ perovskite unit cell), some distortions will occur. As shown in Fig. S\ref{Fig-S4}, the $d_{\text{Pb1-O1}}$ ($d_{\text{Pb2-O2}}$) is larger than $d_{\text{Sr-O1}}$ ($d_{\text{Sr-O2}}$). On the contrary, when a Sr cation is replaced by Pb cation in cubic paraelectric STO supercell ($4 \times 4 \times 4$ perovskite unit cell), the $d_{\text{Sr1-O1}}$ ($d_{\text{Sr2-O2}}$) is smaller than $d_{\text{Pb-O1}}$ ($d_{\text{Pb-O2}}$).  The results indicate that compared with Pb cation, Sr cation has a stronger attracting interaction with its neighbouring O atoms. 

Oxygen octahedral tilts in perovskite usually can attribute to purely steric effect denoted by the tolerance factor $t=(r_O+r_A)/\sqrt{2}(r_O+r_B)$, where $r_O$, $r_A$, and $r_B$ are the O, A-site, and B-site ionic radii, respectively. When the A ion is too small for its site and $t$ is less than unity, the tilts of oxygen octahedra may occur. However, in our PTO-STO heterostructures, the ionic radii for Sr and Pb ions are almost the same (1.26 {\AA} and 1.29 {\AA}, respectively). It suggests that the octahedral tilts might not be caused by different size of Sr and Pb ions. Indeed, for the LP0.5 structure, a titling angle of 0.61$^{\text{o}}$ will satisfy the steric requirement, much smaller than the angle of 2.90$^{\text{o}}$ observed in the fully relaxed structure in DFT calculations. Thus the octrahedra titling could be of covalent bond nature, instead of a steric effect. \cite{Puente2011PRL}

\begin{figure*}[!htb]
\includegraphics[width=120mm]{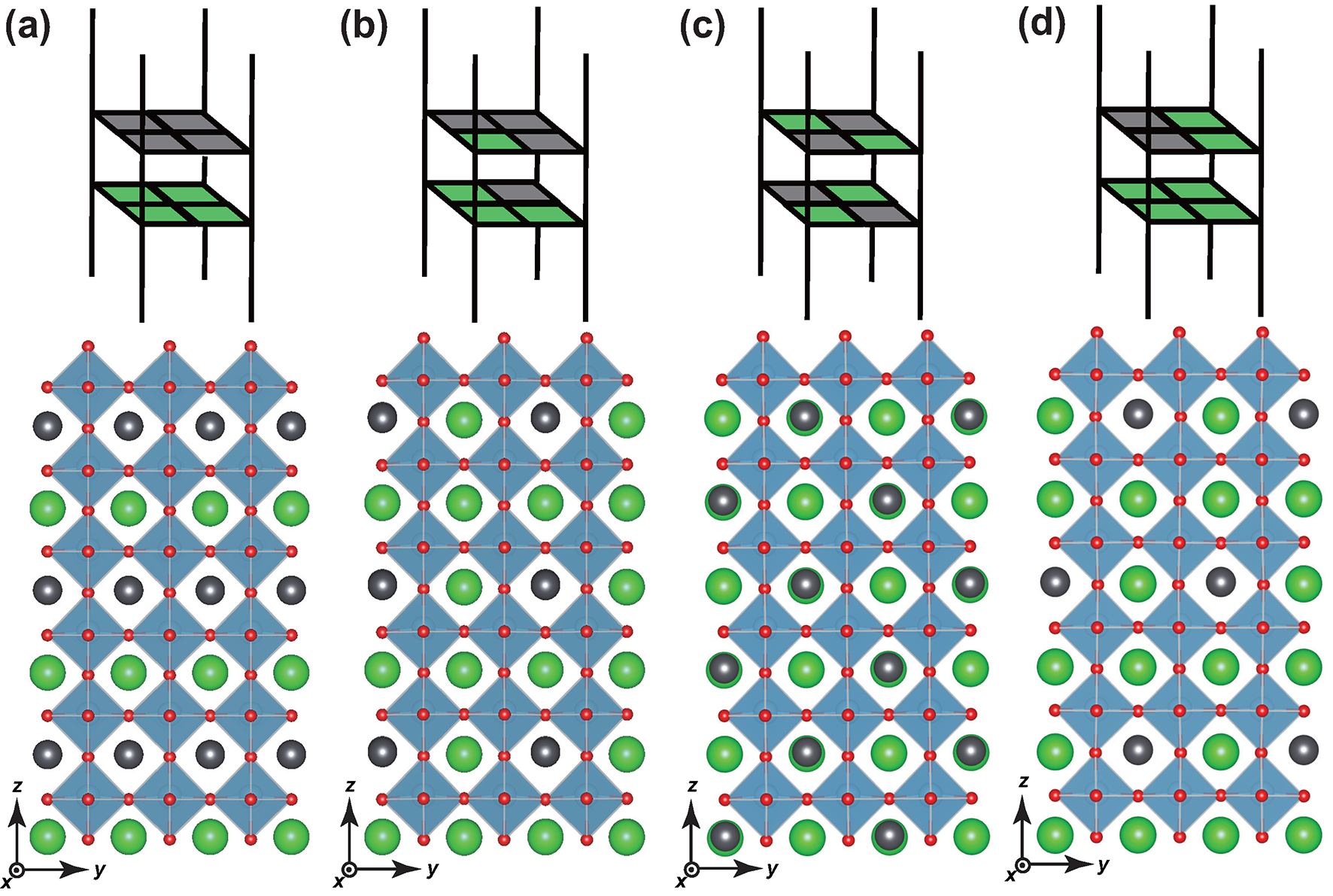}
\caption{\label{Fig-S5} Some cation ordered configurations of PTO-STO at 50\% and 25\% PTO compositions. (a) ultrashort period of PTO-STO (001) superlattice (SL1/1); (b) the 25\% and 75\% cation intermixed layers (n1m1-25/75); (c) the checkered intermixed layers (n1m1-checker) for PTO-STO heterostructures at 50\% PTO composition; (d) the striped intermixed layers of PTO-STO heterostructure at 25\% PTO composition (n3m1-strip).}
\end{figure*}

As discussed in manuscript, the cation ordered patterns of LP0.5 and LP0.25 (Fig. 3 and Fig. S\ref{Fig-S2}) could maximise the AFD distortion modes, such as titling along $x$ or $y$ axis and the octahedral shape distortion in the $yz$ and $xz$ planes. To support this claim, we did a thorough analysis for all of the heterostructures at $x=0.5$ and $x=0.25$ in DFT simulations. Figure S\ref{Fig-S5} shows some typical cation ordered configurations at $x=0.50$ and $x=0.25$. The results for octahedra rotation/tilting angles and Sr-O and Pb-O interatomic distances are summarised in Table S\ref{Tab-S3}. The LP0.5 and LP0.25 configurations do have the largest rotation/titling angles and the shortest Sr-O (longest Pb-O) interatomic distances at their specific concentration.

\begin{table}[!ht]
\caption{\label{Tab-S3} Structural analysis for all PTO-STO heterostructures at 50\% and 25\% PTO composition in DFT. $x$(PTO), $P$, $\theta$, $d_{\text{Sr-O}}$, $d_{\text{Pb-O}}$, $E_{\text{tot}}$ represent concentration of PTO, rotation/tilting angles of octahedral, distances of one O atom and its neighboring Sr/Pb atom in (001) plane, total energy of the relaxed PTO-STO heterostructures.}
\centering
\begin{tabular}{c c c c c c c c c c c c c c c }
\toprule
   Configuration & & {$x$(PTO)} & & {P(C/m$^2$)} & & {$\theta$(degree)} & & {$d_{\text{Sr-O}}$(\AA)} & & {$d_{\text{Pb-O}}$(\AA) } & & {c/a} & & {$E_{\text{tot}}$(eV/unit-cell)} \\ 
\hline
 {SL1/1} & & 50\% & & 0.229 & & 0 & & 2.735 & & - & & 1.013 & & -42,959 \\
 {n1m1-25/75} & & 50\% & & 0.331	& & 0.14 & & 2.699 & & 2.739 & & 1.018 & & -42.968 \\
 {n1m1-checker} & & 50\% & & 0.332 & & 0 & & 2.737 & & 2.742 & & 1.018 & & -42.969 \\
 \textbf{LP0.5} & & \textbf{50\%} & & \textbf{0.391} & & \textbf{2.9} & & \textbf{2.666} & & \textbf{2.817} & & \textbf{1.021} & & \textbf{-42.975} \\
 {SL2/2}	& & 50\%	 & & 0.249 & & 0	& & 2.735 & & - & & 1.013 & & -42.959 \\
 {n2m2-25/75} & & 50\% & & 0.293 & & 0.59 & & 2.735 & & 2.79 & & 1.016 & & -42.964 \\
 {n2m2-checker} & & 50\% & & 0.272 & & 0 & & 2.734 & & 2.739 & & 1.015 & & -42.963
\\
 {n2m2-strip} & & 50\% & & 0.315 & & 1.01 & & 2.693 & & 2.783 & & 1.017 & & -42.966 \\
 {str20} & & 50\% & & 0.348 & & 0.73 & & 2.735 & & 2.805 & & 1.019 & & -42.971 \\
 {str381} & & 50\% & & 0.378 & & 1.73 & & 2.73 & & 2.81 & & 1.02 & & -42.973 \\
 {str410} & & 50\% & & 0.369 & & 2.23 & & 2.714 & & 2.786 & & 1.02 & & -42.971 \\
\hline
 {SL3/1} & & 25\% & & 0.122 & & 0 & & 2.733 & & - & & 1.005 & & -43.537 \\
 {n3m1-strip} & & 25\% & & 0.176 & & 0.77 & & 2.721 & & 2.771 & & 1.007 & & -43.539 \\
 \textbf{{LP0.25}} & & \textbf{25\%} & & \textbf{0.237} & & \textbf{0.87} & & \textbf{2.723} & & \textbf{2.77} & & \textbf{1.009} & & \textbf{-43.54} \\
 {str14} & & 25\% & & 0.166 & & 0 & & 2.733 & & 2.735 & & 1.007 & & -43.538 \\
\toprule

\end{tabular}
\end{table}

\begin{figure*}[!htb]
\includegraphics[width=80mm]{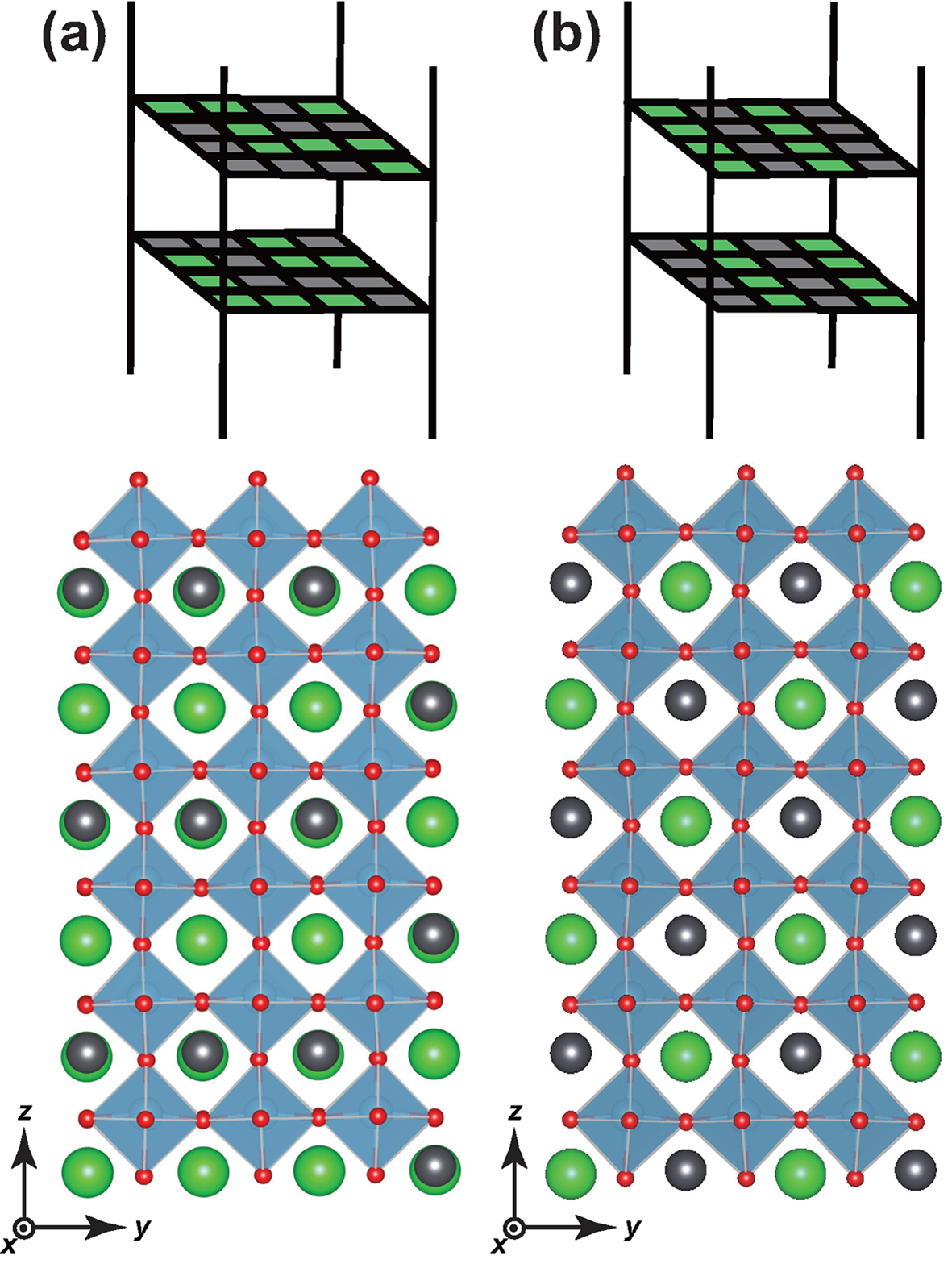}
\caption{\label{Fig-S6} Crystal structure of (a) the "381" configuration and (b) the LP0.5 configuration. The two top figures depict the distributions of Pb (grey grids) and Sr (green grids) cations.}
\end{figure*}

Figure S\ref{Fig-S6}(a) shows a specific heterostructure "381" (named as str381) whose cation ordering pattern resembles that of LP0.5. We artificially turned off its AFD distortion modes in DFT calculations (meanwhile allowing the FE polarisation to fully develop). It is clear that in this intermixing case, str381 has a lower total energy and a higher $P$ value than those of LP0.5 (Table S\ref{Tab-S4}). If the AFD distortion modes are allowed in both configurations, the LP0.5 has a lower total energy value and a higher $P$ value. It is clear that the AFD modes in LP0.5 give rise to a larger polarisation enhancement. This should come from the cation ordering pattern of LP0.5. The check-board cation ordering pattern in the $yz$ plane [Fig. 3(a)] ensures the two Sr cations (neighbouring a Pb cation) can work collaboratively to maximise the $\Gamma_4^+$ and $\Gamma_1^+$ modes, as shown in Table S\ref{Tab-S3}. If a small variation was introduced in the ordered pattern of LP0.5, such as str381 (\emph{i.e.}, two Pb and two Sr cations swap their positions), a weaker collaborative effect is observed, resulting in a smaller tilting angle and Ti-O bending angle (Table S\ref{Tab-S4}). Thanks to the AFD/FE coupling, the LP0.5 has a larger $P$ value than the str381.

\begin{table}[!ht]
\caption{\label{Tab-S4} Comparison of ferroelectric polarisation $P$, octahedra titling angle, c/a ratio, and total energy results between str381 and LP0.5. For the intermixing cases, AFD distortion modes are restricted meanwhile the FE polarisation modes are allowed in DFT calculations.}
\centering
\begin{tabular}{c p{0.5cm} c p{0.5cm} c p{0.5cm} c p{0.5cm} c }
\toprule
   & & {str381-intermixing} & & {LP0.5-intermixing} & & {str381} & & {LP0.5} \\
\hline
 {polarization (C/m$^2$)}   & & 0.360   & & 0.355 & & 0.378 & & 0.391 \\
 {tilting angle (degree)}   & & -       & & -     & & 1.50   & & 2.90  \\
 {c/a ratio}                & & 1.018   & & 1.019 & & 1.020 & & 1.021  \\
 {Energy (eV/Unit-cell)}    & & -42.964 & & -42.963 & & -42.973 & & -42.975  \\
\toprule
\end{tabular}
\end{table}



\pagebreak

\pagebreak
\section{APPENDIX}

\begin{longtable}{ c c c c c c c c}
\caption{\label{Tab-S5}A description of crystal structure of LP0.5 and LP0.25.}\\
\hline\hline
   {Structure} & {Space} & {Wyckoff} & {x} & {y} & {z} & {$a/b/c$}    & {$\alpha/\beta/\gamma$}\\
        {ID}       &  {group}                      & {position}  &      &      &      & (\text{\AA}) & \\   
\hline
LP0.5 & 6 & 1a (Sr) & 0.00000 & 0.00000 & 0.02150 & 7.7280 & 90.000 \\
           &    & 1b (Sr) & -0.00002 & 0.50000 & 0.02152  & 7.7280 & 90.000 \\
           &    & 1a (Sr) & 0.50000 & 0.00000 & 0.52148  & 7.8826 & 90.000 \\
           &    & 1b (Sr) & 0.50000 & 0.50000 & 0.52151  &  &  \\
           &    & 1a (Pb) & 0.50000 & 0.00000 & 0.03725  &  &  \\
           &    & 1b (Pb) & 0.49999 & 0.50000 & 0.03721  &  &  \\
           &    & 1a (Pb) & 0.00001 & 0.00000 & 0.53725  &  &  \\         
           &    & 1b (Pb) & 0.00000 & 0.50000 & 0.53721  &  &  \\         
           &    & 2c  (Ti) & 0.24703 & 0.75005 & 0.27167  &  &  \\
           &    & 2c  (Ti) & 0.75299 & 0.75005 & 0.27167 &  & \\
           &    & 2c  (Ti) & 0.25298 & 0.75004 & 0.77168  &  &  \\
           &    & 2c  (Ti) & 0.74701 & 0.75005 & 0.77167 &  & \\
           &    &  2c (O) & 0.23618 & 0.74992 & 0.00274  &  &  \\
	  & &  2c (O) & 0.00000 & 0.74762 & 0.25371  &  &  \\ 
& &  1a (O) & 0.25370 & 0.00000 & 0.24808  &  &  \\ 
& &   1b (O) & 0.25846 & 0.50000 & 0.24853  &  &  \\
& &  2c (O) & 0.76383 & 0.74995 & 0.00274  &  &  \\
& & 2c (O) & 0.50000 & 0.75245 & 0.24119  &  &  \\ 
& & 1a (O) & 0.74632 & 0.00000 & 0.24810  &  &  \\
& &  1b (O) & 0.74156 & 0.50000 & 0.24850  &  &  \\
& &  2c (O) & 0.26383 & 0.74995 & 0.50274  &  &  \\
& &  2c (O) & 0.00000 & 0.75244 & 0.74120  &  &  \\
& &  1a (O) & 0.24630 & 0.00000 & 0.74810  &  &  \\
& &  1b (O) & 0.24155 & 0.50000 & 0.74851  &  &  \\
& &  2c (O) & 0.73617 & 0.74993 & 0.50273  &  &  \\
& &  2c (O) & 0.50000 & 0.74762 & 0.75371  &  &  \\
& &  1a (O) & 0.75369 & 0.00000 & 0.74808  &  &  \\
& &  1b (O) & 0.75844 & 0.50000 & 0.74853  &  &  \\
\hline        
LP0.25 & 8 & 2a (Sr) & 0.51021 & 0.00000 & -0.00008 & 7.7955 & 90.000 \\
          &    & 4b (Sr) & 0.76010 & 0.25001 & 0.49993  & 10.929 & 125.50 \\
          &    & 2a  (Pb) & 0.02577 & 0.00000 & -0.00023  & 6.7122 & 90.000 \\
          &    & 2a  (Ti) & 0.00991 & 0.00000 & 0.49693 &  & \\
          &    & 2a  (Ti) & 0.51271 & 0.00000 & 0.50293 &  & \\
          &    & 4b  (Ti) & 0.76133 & 0.24850 & -0.00007 &  & \\          
          &    & 2a  (O) & 0.74676 & 0.00000 & 0.49464  &  & \\
          &    & 4b  (O) & 0.62234 & 0.37548 & 0.75096  &  &  \\
          &    & 2a  (O) & 0.25209 & 0.00000 & 0.50552 &  & \\
          &    & 4b  (O) & 0.11842 & 0.37299 & 0.74604 &  &  \\
          &    & 4b  (O) & 0.49939 & 0.24728 & 0.00003 &  &  \\         
          &    & 4b  (O) & 0.87243 & 0.12702 & 0.25411 &  &  \\
          &    & 4b  (O) & 0.87146 & 0.37544 & 0.24913 &  &  \\
\hline\hline
\end{longtable}
